\documentclass[rnote]{aa} 
\usepackage{graphicx}
\usepackage{txfonts}
\usepackage{natbib}
\bibpunct{(}{)}{;}{a}{}{,}

\begin{document}
\title{Influence of major mergers on the radio emission of elliptical galaxies}

\author{Y.~H. Wen,  Z.~L. Wen, J.~L. Han \and L.~G. Hou}
\authorrunning{Wen et al.}
\titlerunning{Influence of major mergers on the radio emission of elliptical galaxies}
\institute{National Astronomical Observatories, CAS, 
Jia-20 DaTun Road, Chaoyang District, Beijing 100012, China.
zhonglue@nao.cas.cn}

\date{Received; accepted ...}

\abstract
{} 
{We investigate the influence of major mergers on the 
radio emission of elliptical galaxies.}
{We use a complete sample of close pairs, which contains 
475 merging and 1828 non-merging paired elliptical galaxies of 
$M_{r}<$-21.5 selected from the Sloan Digital Sky Survey. 
In addition, a control sample of 2000 isolated 
field galaxies is used for comparison. We cross-identify the optical galaxies 
with the radio surveys of FIRST and NVSS.}
{We find that the radio fraction of merging paired galaxies is about 6\%, 
which is slightly higher than the 5\% obtained for non-merging paired galaxies, 
although these values are consistent with each other owing to the large 
uncertainty caused by the limited sample. The radio fraction is twice as 
that of isolated galaxies, which is less than 3\%.}
{Radio emission of elliptical galaxies is only slightly affected 
by major mergers, but predominantly depends on their optical luminosities.
Therefore, merging is not important in triggering the radio emission of 
elliptical galaxies.}


\maketitle

\section{Introduction} 
\label{intro}
The interactions of galaxies and merging processes are very common in
the universe, and they affect the properties of galaxies, for instance
their morphology \citep{wr78}, star formation
\citep{sse+88,bgk00,lta+03,rbs+09}, and the activity of their nuclei
\citep{kkh+85,sts+07,jmp+08}.

Previously, there have been a number of investigations of the radio
emission related to interactions and the merging
process. \citet{sto78} found that both spiral and elliptical galaxies
in close pairs are more likely to be radio sources, twice as that of
the widely separated galaxies. For star-forming galaxies, radio
emission is enhanced by star formation triggered by galaxy
interactions \citep{hum81,ccg+82,ap84,hec83,gbm+90}.
Radio galaxies with strong optical emission lines tend to have
interaction features in optical images, which are probably produced by
the merging of gas-rich galaxies
\citep{hsb+86,mvs+87,gtp+06,thg+11,rbt+11}.
\citet{wbk+87} and \citet{tdn+09} showed that interactions play only
a minor role on the nuclear activity of elliptical galaxies. The radio
detection rates of elliptical galaxies is found to be directly related
to the optical luminosity \citep{cff89}, rather than other properties of
galaxies, such as ellipticity and shape of isophotes \citep{gff+00}.

Previous investigations of the radio emission caused by physical
interactions or mergers of gas-poor elliptical galaxies were affected
by the small size of samples and the difficulty in the merger
identification. Only visual pairs, which may or may not be mergers,
were used in all kinds of studies. The merging galaxies clearly host
stronger interactions than the non-merging ones. The interaction
features of elliptical galaxies are very weak and far more difficult
to identify than those of spiral galaxies.
In this paper, we verify whether any enhancement in radio emission is
caused by the major dry mergers of elliptical galaxies using a large
volume-limited complete sample of pairs that was identified in the
Sloan Digital Sky Survey data release 6 (SDSS DR6).

\section{Radio identification of elliptical galaxies}
\label{crossid}

\subsection{Samples of elliptical galaxies}

On the basis of the SDSS DR6, \citet{wlh09} obtained a large volume-complete
sample of 1209 pairs of luminous elliptical galaxies by applying the
following criteria:

1. Each galaxy has an extinction-corrected Petrosian magnitude of
$13.5 < r < 17.5$ and a rest-frame absolute magnitude of $M_r < -21.5$.

2. Rest-frame colors of galaxies satisfy $(u - r) > 2.2$ and $(g - r) >
0.7$.

3. Close pairs have a projected separation of 7~kpc $< r_p <$ 50~kpc
and a redshift of $z < 0.12$.

\citet{wlh09} fitted each pair of galaxies with a smooth
surface-brightness profile. Residual images of close pairs were
obtained by subtracting the smoothed profiles from the original
observational images.  Interaction features (e.g., tails, bridges and
plumes) in these pairs were identified based on an image
analysis. Galaxies in 249 close pairs which display distinct
interaction features in the residual images are classified as merging
galaxies. Galaxies in 960 close pairs which do not have distinct
interaction features are classified as non-merging ones. In this work,
we only consider the paired galaxies with a fitting magnitude of
$M_r<-21.5$, which corresponds to 475 merging galaxies and 1828
non-merging galaxies.

To access the influence of interactions on the radio emission of elliptical
galaxies, we define a control sample of 2000 isolated field elliptical
galaxies, by selecting these galaxies without bright nearby companions
within a radius of 500~$h^{-1}$~kpc and a velocity of $\pm$700~km
s$^{-1}$. This sample of isolated galaxies has similar
distributions of color, luminosity, and redshift as those of the paired
galaxies.

\begin{figure}
\centering
\includegraphics[width=80mm,viewport=20 70 525 425]{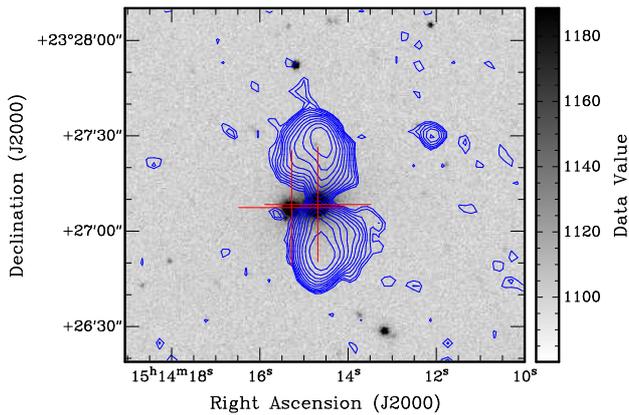}
\caption{Examples of overlapping images of SDSS $r$-band optical
  image in grey with the contours for radio emission from
  the FIRST data. The crosses indicate the locations of the optical
  galaxies. 
\label{example}}
\end{figure}

\subsection{Radio emission from ellipticals}

As done by \citet{bkh+05a}, we verify our detection of radio emission
from those merging, non-merging, and isolated elliptical galaxies
using the data from the FIRST and NVSS surveys. The FIRST survey was
carried out at a wavelength of 20~cm, which covers 9900 deg$^2$ of the
northern sky \citep{bwh95} and overlaps with most of the sky area of
the SDSS DR6. The FIRST survey has a resolution of 5 arcsec and a flux
density limit of 1 mJy for point sources.
Radio galaxies often have widely separated double lobes or many
resolved components. The FIRST survey was often able to resolve
galaxies into several components because of its high resolution, so
that the source flux usually refers to that of each component.  To get
the total flux of a radio source, we supplementarily use the data from
the NVSS survey, which has a lower resolution of 45 arcsec at 1.4~GHz
and covers the entire northern sky of declination $> -$40$^{\,\rm o}$
\citep{ccg+98}. The NVSS data is complete to a flux limit of 2.5 mJy
for point sources. Not all the close pairs identified from the SDSS
DR6 are located at the sky area of the FIRST survey. Forty-five pairs
are outside the FIRST survey area but within the NVSS region. All
isolated galaxies are selected from the common area of the FIRST and
NVSS surveys.

We superimpose the NVSS contour maps on the SDSS images of pairs. We
identify the radio emission from galaxies based on positional
coincidence. Owing to the low resolution of NVSS, radio sources within
10 arcsec of the SDSS paired galaxies are considered good
matches. Galaxies usually have radio jets or lobes that extend to
hundreds of kpc. Some have weak radio emission at the optical
position. Visual inspection of the superimposed optical and radio
images enables us to readily identify a counterpart galaxy even if
jet-like sources are outside the optical image. We also superimpose
the FIRST contour maps on the SDSS images. The FIRST sources within 2
arcsec of the SDSS galaxies are considered as radio
detections. Figure~\ref{example} shows an example of radio lobes
originating from a galaxy by visual inspection. The same procedures
are applied to 2000 isolated field galaxies.

\begin{figure}
\centering 
\includegraphics[width=75mm]{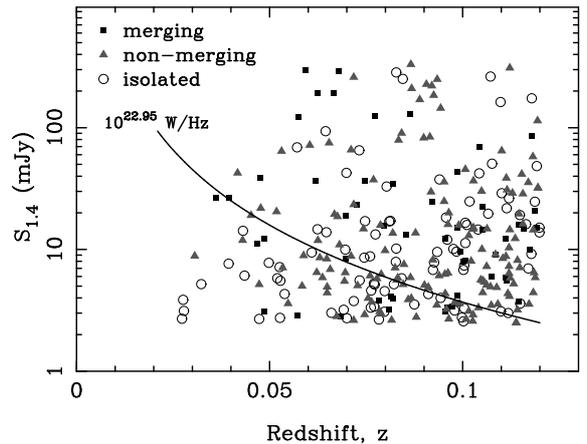}
\caption{Radio flux of galaxies against redshift.
  The solid line shows the radio power cutoff of $P_{1.4}= 10^{22.95}$
  W~Hz$^{-1}$ for complete samples.
\label{fluxz}}
\end{figure}

\begin{figure*}[th!]
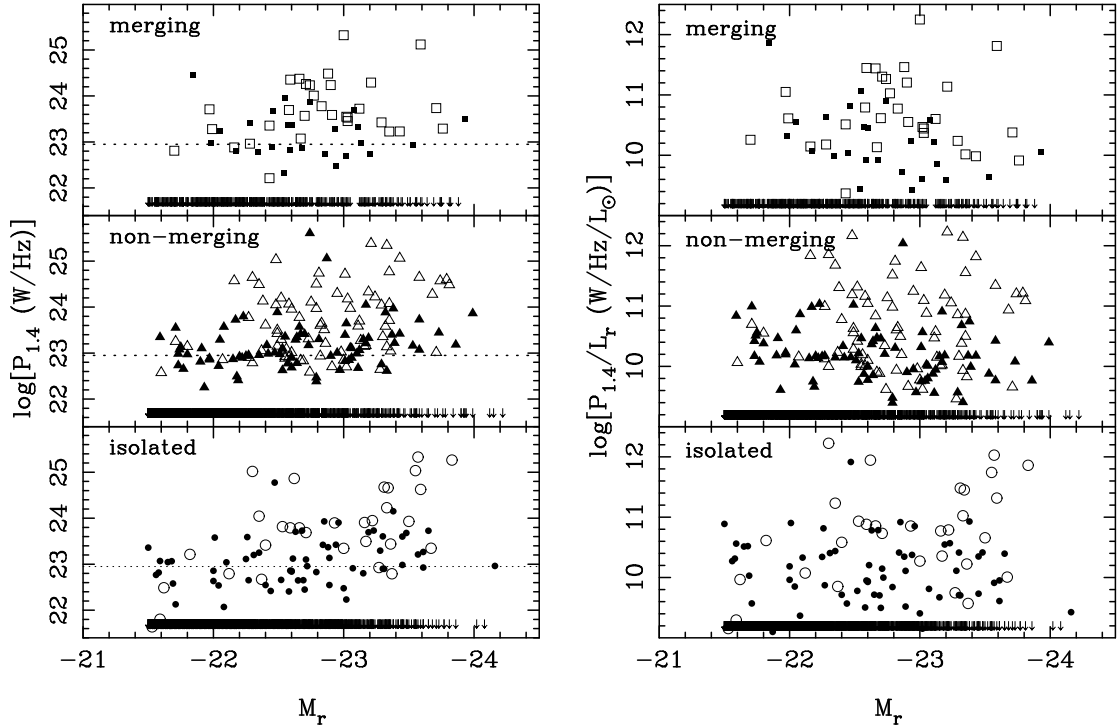

\centering 
\includegraphics[width=70mm]{f3a.ps}\hspace{5mm}
\includegraphics[width=70mm]{f3b.ps}
\caption{Left panels: radio power against the absolute magnitude for
  galaxies in three samples. Galaxies with jet-like radio sources are
  denoted as open symbols, and those with compact sources are denoted
  as filled symbols. The dotted line represents the cutoff for a
  complete sample of radio objects. The down-wards arrows are the
  upper limit of the undetected galaxies in radio. Right panels: the
  same but for radio-to-luminosity ratio against the absolute
  magnitude for three samples.
\label{pwmag}}
\end{figure*}

To be able to access most effectively the radio emission properties of
merging or non-merging galaxies, we should set a radio power cutoff
rather than use the flux limit of the radio surveys. Because the close
pairs are a volume-limited complete sample at $z<0.12$. The radio
power of an elliptical galaxy is calculated to be $P_{1.4} = 4\pi
D_l^2S_{1.4}$, where $S_{1.4}$ refers to NVSS radio flux in 1.4~GHz and 
$D_l$ is luminosity distance from the galaxy. The radio power cutoff
is set to be $P_{1.4}=10^{22.95}$ W~Hz$^{-1}$, above which radio
sources of $z<0.12$ for the volume-limited complete sample have a flux
limit of 2.5 mJy. If we consider the radio sources with a flux limit
of 5 mJy, the sample is complete for the radio sources of
$P_{1.4}>10^{23.25}$ W~Hz$^{-1}$. Figure~\ref{fluxz} shows the radio
flux against redshift for galaxies, with a flux limit of 2.5 mJy.

Figure~\ref{pwmag} shows the radio power and the ratio of radio power
to optical luminosity (i.e., radio-to-luminosity ratio) against the
optical luminosity for the three samples. The radio emission tends to
be stronger for galaxies with higher optical luminosities, which is very
consistent with the conclusions obtained by \citet{cff89} and
\citet{bkh+05b}. The radio emission of galaxies have two morphology types,
compact sources, and jet-like sources. Galaxies with jet-like radio
emission tend to have a stronger radio power than those
with compact radio sources.

\subsection{Fraction of radio detections and interactions}

The fraction of radio detections is defined as the number of galaxies
with radio emission above a threshold divided by the total number of
galaxies in the sample. If we take a radio power of
$P_{1.4}=10^{22.95}$ W~Hz$^{-1}$ as the threshold, we get 154 radio
detections of galaxies in close pairs, of which 42 are merging
galaxies and 112 are non-merging galaxies. Similarly, we find 59 radio
detections out of 2000 isolated field elliptical galaxies. The
fractions of radio detections are 8.8\%$\pm$1.4\% for the merging
galaxies, 6.1\%$\pm$0.6\% for the non-merging galaxies, and
3.0\%$\pm$0.4\% for the isolated galaxies. If we take the threshold of
$P_{1.4}=10^{23.25}$ W~Hz$^{-1}$, the fractions become
7.4\%$\pm$1.2\%, 4.2\%$\pm$0.5\%, and 2.3\%$\pm$0.3\% for the three
samples, respectively.

\begin{figure}[t!]
\centering 
\includegraphics[width=75mm]{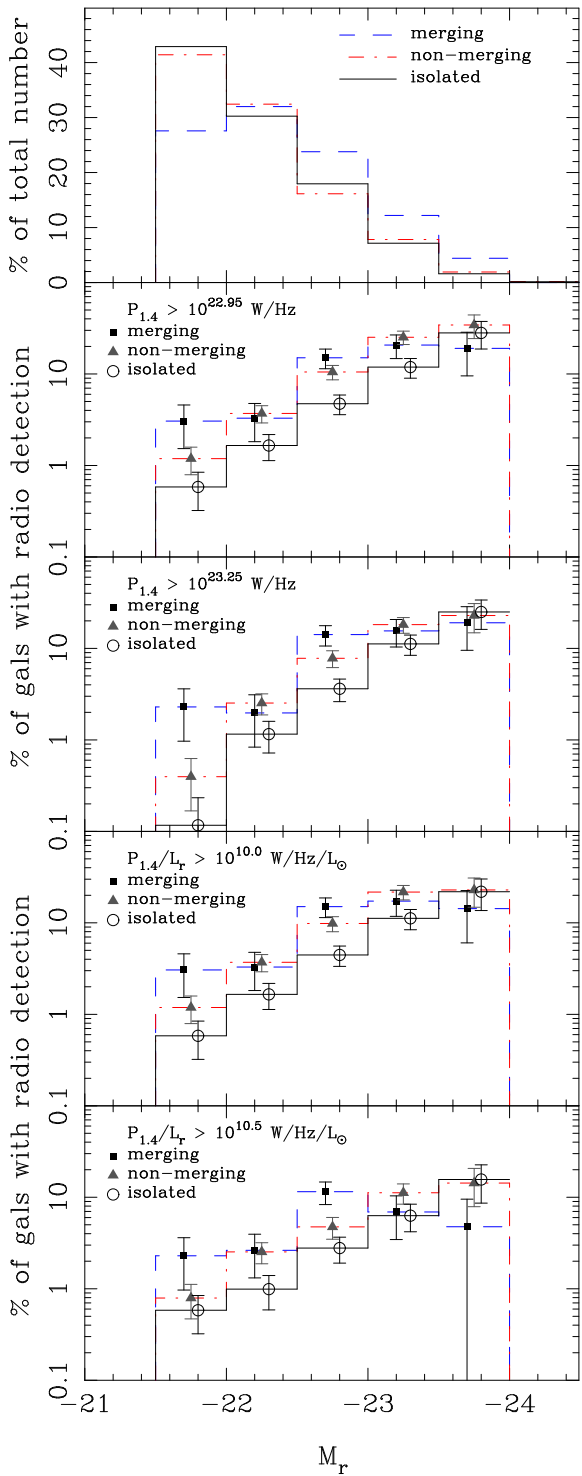}
\caption{Top panel: distributions of absolute magnitude for three
  samples.  Second and third panels: radio fraction after applying 
  thresholds for a radio detection in terms of radio power. Fourth and fifth
  panels: radio fraction after applying the thresholds of radio-to-luminosity
  ratio.
\label{ratemag}}
\end{figure}

Figure~\ref{ratemag} shows the dependence of the fraction of radio
detections on optical absolute magnitude for both radio power
thresholds (the second and third panels). More luminous galaxies are
clearly more likely to be radio sources for all three samples. Owing
to the dependence on luminosity, i.e., absolute magnitude, we should
be very careful to compare the radio fractions of these samples with
different absolute magnitude distributions (see the top panel of
Figure~\ref{ratemag}). To minimize the luminosity effect, we calculate
the fractions (or the numbers) of radio detections within five
absolute magnitude bins. We then normalize the percentage of total
galaxy number within each absolute magnitude bin for merging and
non-merging galaxies to ensure that they are the same as that of
isolated galaxies (solid line of Figure~\ref{ratemag}) by multiplying
by a factor. Both the total numbers of galaxies and radio detections
within each absolute magnitude bin are then multiplied by the
factor. Finally, all these numbers are added together within the
absolute magnitude range of $-24.0<M_r<-21.5$ to give the normalized
total numbers of galaxies and radio detections. We finally get the
normalized fraction for the three samples for comparison.
As listed in Table~\ref{tab}, the radio fraction of merging galaxies
is about 6\%, which is slightly higher than the value of about 5\% for
non-merging galaxies, but not significantly so owing to the error
bars. These values are twice as large as that of isolated galaxies,
which is less than 3\%.

It is necessary to test whether the above results depend on the
definition of the radio detection threshold. \citet{bkh+05b} suggests
that the radio-to-luminosity ratio is a more appropriate definition of
radio detection. As shown in Figure~\ref{ratemag} (the fourth and
fifth panels), the fractions of radio detections also strongly depend
on absolute magnitude when the radio detection is defined in terms of
the radio-to-luminosity ratio.
Again, after normalizing the distribution of absolute magnitude, we
get the fraction of radio detections for three samples by applying the
thresholds of the radio-to-luminosity ratio, as listed in
Table~\ref{tab}. The fraction of merging galaxies is 1\% higher than
but still consistent with that of non-merging galaxies. The fraction
of radio detections of merging or non-merging galaxies is almost
double that of isolated galaxies, but just with a significance of
$\sim$$\,2\sigma$.

\begin{table}[t!]
\caption{Fraction of radio detections of galaxies with distribution of
  absolute magnitude normalized to that of isolated galaxies.}
\label{tab}      
\centering          
\begin{tabular}{l c c c}
\hline
  Threshold of & \multicolumn{3}{c}{Radio fraction (\%)} \\
radio detection  & Merging & Non-merging & Isolated  \\
\hline                    
$P_{1.4}>10^{22.95}$ W/Hz & 6.8$\pm$1.5 & 5.8$\pm$0.6 & 2.9$\pm$0.4\\
$P_{1.4}>10^{23.25}$ W/Hz & 5.5$\pm$1.3 & 4.0$\pm$0.5 & 2.3$\pm$0.3\\
$P_{1.4}/L_r>10^{10.0}$ W/Hz/L$_{\odot}$ & 6.5$\pm$1.5 & 5.3$\pm$0.5 & 2.7$\pm$0.4\\
$P_{1.4}/L_r>10^{10.5}$ W/Hz/L$_{\odot}$ & 4.4$\pm$1.2 & 3.0$\pm$0.4 & 1.8$\pm$0.3\\
\hline                  

\end{tabular}
\end{table}

\begin{figure}
\centering 
\includegraphics[width=75mm]{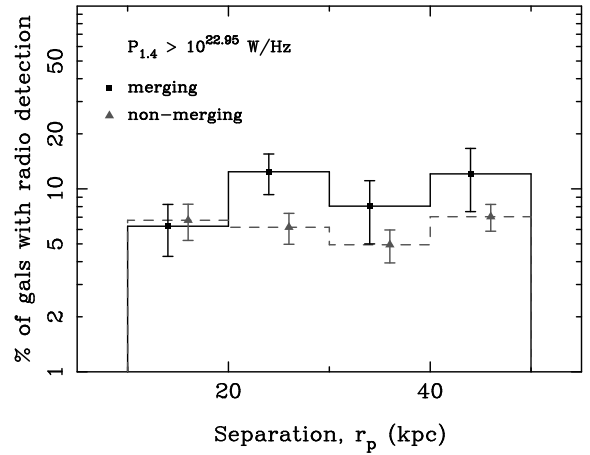}
\caption{Fraction of radio detection against the galaxy separation for
  close pairs.
\label{raterp}}
\end{figure}

Our results suggest that the radio emission of elliptical galaxies is
slightly affected by galaxy interactions. The radio emission
predominantly depends on galaxy luminosity. This conclusion is
consistent with previous investigations
\citep{ape+77,hke83,hec83b,fgg+84,wbk+87,cff89,tdn+09,bjf+11}.  We
also check whether the radio detection depends on the separation of
close pairs. As shown in Figure~\ref{raterp}, there is no significant
dependence of radio fraction of paired galaxies on separation. The
merging ellipticals are slightly more likely to be radio sources than
non-merging ellipticals for all separations, but the uncertainties are
large owing to the small number of merging galaxies.

\section{Conclusions}

We have compared the radio emission of 475 merging galaxies and 1828
non-merging galaxies in close pairs and a sample of 2000 isolated
field galaxies. A cross-identification of these galaxies with the
FIRST and NVSS radio data shows that 154 galaxies have radio
detections, of which 42 are merging galaxies, 112 are non-merging
galaxies. In addition, 59 isolated galaxies have radio detections. The
fraction of radio detections depends strongly on optical
luminosity. The fraction of radio detections of merging galaxies is
about 6\%, which is about 1\% higher than that of non-merging galaxies
and twice as that of isolated galaxies. Our results indicate that
merging is not important in triggering the radio emission of
elliptical galaxies. This is consistent with the conclusions of
\citet{stu+11} and \citet{kfm+12}.

\begin{acknowledgements}
We thank the referee for valuable comments that helped to improve
the paper. We are grateful to Jun Xu, Wei Cai, and Dr. YuanJie Du for
helps. The authors are supported by the National Natural
Science Foundation of China (10833003 and 11103032).
\end{acknowledgements}

\end{document}